# Service-Oriented Architecture in Industrial Automation Systems - The case of IEC 61499: A Review

Kleanthis Thramboulidis, *Member, IEEE*

*Abstract—* In the paper by W. Dai *et al.* (*IEEE Trans. On Industrial Informatics*, vol. 11, no. 3, pp. 771-781, June 2015), a formal model is proposed for the application of SOA in the distributed automation domain in order to achieve flexible automation systems. A SOA-based execution environment architecture based on the IEC 61499 Function Block model is proposed and a case study is used to demonstrate dynamic reconfiguration. In this letter, a review of the literature related to the use of SOA in Industrial Automation Systems is given to set up a context for the discussion of the proposed in the above paper SOA IEC 61499 formal model. The presented, in the above paper, formal model and the execution environment architecture are discussed towards a better understanding of the potentials for the exploitation of the SOA paradigm in the industrial automation domain.

## I. Introduction

Authors in the first issue of IEEE Transactions on Industrial Informatics, ten years ago, described opportunities and challenges in using the service oriented architecture in manufacturing [1]. Since then several research articles published in the same journal reporting successful or promising results regarding the exploitation of the SOA paradigm in the industrial automation system (IAS) domain, e.g., [2][3][4]. Similar results have been published in other journals too, e.g. [5].

In the last issue of the journal, i.e., June 2015, authors present in [7], a formal model for the application of SOA in the distributed automation domain in order to achieve flexibility. They adopt the IEC 61499 standard [25] instead of the widely used in industry IEC 61131 [26], for several reasons they present in the paper. They also describe an execution environment based on the proposed formal model and demonstrate the flexibility of the proposed approach by a scenario for dynamic reconfiguration.

In this letter the proposed approach is discussed in the context of both the SOA paradigm and the IEC 61499 Function Block model, in an attempt to identify advantages and disadvantages, and its potential for exploitation.

The remainder of this letter is organized as follows. Section II discusses published work regarding the exploitation of SOA in the industrial automation domain, in order to set up a framework for the discussion. Section III discuss the SOA based IEC 61499 model presented in [7]. Section IV comments on the SOA-based execution environment architecture and the run-time reconfiguration. Finally, Section V concludes this letter.

## II. SOA in Industrial Automation Systems

SOA was considered for several years as one of the hottest subjects in the IT community. This has been changed the last few years when IoT has replaced this buzzword. As expected, SOA has attracted the attention of researchers in the industrial automation domain. Several research groups presented their work towards the exploitation of the SOA paradigm in IASs.

Authors in [1] outline opportunities and challenges in using the service oriented architecture in the manufacturing community. They claim that web services technology constitutes the preferred implementation vehicle for service-oriented architectures and they discuss the extension of the SOA paradigm into the device space that will allow to seamlessly integrate device-level networks with enterprise-level networks. Authors capture the disadvantages of UPnP (Universal Plug and Play) initiative, already used in industry, in comparison with web services. The key concepts of the SIRENA project [13], which was is part of the ITEA initiative, are described.

SIRENA has played a pioneering role by applying the SOA paradigm to communications and interworking between components at the device level and its results were used as a foundation for both SODA [14] and SOCRADES [15] projects. SODA exploited the framework of SIRENA and defined it in a platform, language and network neutral way, applicable to a wide variety of networked devices in several domains among which IASs. It has also promoted a Devices Profile for Web Services (DPWS) as an OASIS standard [30] and delivered different implementations. An implementation of the DPWS specification based on the J2ME CDC platform was developed by the SOA4D (Service-Oriented Architecture for Devices) [16] open-source initiative for exploiting and adapting SOAP [28] and Web services to the specific constraints of embedded devices. SOCRADES proposes the use of SOA as Web services in such a way that it results to a unifying application-level communication mean across the various levels of the enterprise pyramid down to the device level, for the devices to expose selected functionality to be used by the layers above.

In [2], authors present a SOA based framework for Industrial Automation enhanced with real-time capabilities. A key characteristic of the proposed framework is that it allows for negotiation of the QoSs requested by clients from web services, and provides temporal encapsulation of individual activities. This allows to perform an *a priori* analysis of the temporal behavior of each service, and to avoid unwanted interference among them. Authors evaluate current implementations of CORBA [29], such as TAO, that satisfy the requirements of embedded real time system, regarding the requirements they have defined, and argue on the selection of SOAP instead of CORBA as basis for their framework. CORBA is one of the first implementations of the SOA concept for distributed systems. Authors in [5] present a CORBA Component Model (CCM) implementation of the IEC 61499 run-time environment that exports its services to the environment through the CORBA bus. TAO [6], a real-time ORB that implements the real-time CORBA 1.x is utilized.

Authors in [3] present an approach to exploit SOAP in the domain of Evolvable Production Systems. Their approach was inspired by the Devices Profile for Web Services (DPWS) specification, which was extended to address the specific needs of this domain. Programmable Logic Controllers (PLCs) were used as devices. This work is highly related with SODA and SOCRADES projects.

In [4], authors evaluate the performance of PLC-to-PLC communications based on HTTP and compare it to Modbus TCP. The motivation for this work is the appearance, during past years, in the market of various PLCs with embedded HTTP servers. These PLCs may be used in collaboration with PLCs that acts as the HTTP clients, to allow the integration of control systems with soft real-time constraints. Authors claim that while SOA's suitability is proven in IT systems, it has not been adopted yet in commercial PLCs, and thus cannot be considered as a solution for integration with already deployed control systems. They come up with results that indicate that Modbus TCP protocol is significantly better than HTTP and they attribute this result mainly to the relatively low performance of PLC application code executing complex string processing required by the HTTP protocol. They also mention that HTTP performs well enough to meet specified soft real-time constraints of the sample Networked Control System (NCS). The 99.9% of measured HTTP data exchanges are completed in less than 700 ms which makes, as claimed by the authors, the HTTP communications an alternative that is worth evaluating for soft real-time NCS.

Authors in [17] describe an open source SOA architecture for IAS that is composed of three layers. The first layer, which is used to model the information from the device level, is constructed as a set of OPC servers. The second layer, which is used as a link to the third layer is composed of basic and complex services. The third layer, which is named constraint satisfaction problem (CSP) layer, is used for computation of production plans. They demonstrated and evaluated the proposed framework on Apache CFX with SOAP and Jersey, that is an implementation of the JAX-RS, i.e., the Java API for RESTFul web services, and the java based framework Apache River. The use of the SOA paradigm is adopted outside of the device boundary, thus this approach does not consider determinism and real-time deadlines imposed by device level requirements.

Authors in [18] present the application of SOA in building automation systems. The presented approach utilizes the DPWS profile, ontologies for representing semantic data, and a composition plan description language to describe context-based composite services in form of composition plans. They claim that SOAP and WSDL is the most popular implementation of SOA which is gaining increasing market penetration. Authors evaluate four different implementations of the DPWS, two based on C and two based on Java. Authors present evaluation results regarding the feasibility and scalability of the proposed system and specifically the performance overhead of the service selection and service execution processes (composition time). Composition time has been measured as 1000 msec for 500 devices on an Intel processor with 2.6-GHz and 6-GB RAM.

Semantic web services are utilized by authors in [19] to present an approach for managing production processes. Based on this approach devices expose web service interfaces formulated in OWL-S through which they can be controlled. Even though authors claim that the exposed web services interface of the device is used for controlling the device and thus inserting the framework's overhead in the control loop of the plant, performance evaluation is not given with the argument that it is difficult to find similar semantic web service monitoring and composition approaches against which to compare with.

SOAP has been defined as a lightweight protocol intended for exchanging structured information in a decentralized, distributed environment [9]. Authors in [10] investigate CORBA and SOAP as communication mechanisms to interconnect different systems and argue that "it turns out that a direct and naive use of SOAP would result in a response time degradation of a factor 400 compared to CORBA." Since then web services technology had further improved regarding XML parsers but not to the level of considering it as a glue to interconnect constituent components of a controller running on the same device. Even the use of HTTP at the device level is introducing performance overhead that allows the approach to be considered only for soft real-time NCS [3]. SOAP is also not the preferred technology for the IoT where the REST architectural model is considered as the dominating one [11]. SOA is an enabling technology for IoT which is becoming increasingly popular as claimed in [12]. However, is should be noted that the four-layer SOA presented in [12] for the IoT, places the service layer on top of the Network one that is on top of the sensing layer [12, Fig. 4] for which the universal unique identifier (UUID) is considered as key characteristic of IoT to enable the identification and use of provided by devices services. Authors claim that a device with UUID can be easily identified and retrieved. Application API and interface are captured at the interface layer along with Contracts. It is also worthwhile to note that the Service bus is on top of the Business logic.

SOA based products have already appeared in the industrial systems market in the context of Industry 4.0. For example, TwinCAT from Beckhoff combines IEC 61131-3-based SOA services with OPC UA interoperability [20].

## III. THE FORMAL SOA IEC 61499 FUNCTION BLOCK MODEL

SOA was introduced as an approach to design a software system to provide services either to end-user applications or other services distributed in a network, via published and discoverable interfaces [8]. Authors in [7, Sec. 1] admit that SOA has been introduced to facilitate the creation of distributed networked computer systems. However, the formal model they propose utilizes SOA for the integration of software modules that constitute a controller running on a single computation node (device). Based on Definition 4, Function Block Instances (FBIs) are service providers since each input event of an FBI is considered as a provided service. A basic Function Block (FB) is considered to provide atomic services (Definition 2). Moreover, based on Definition 5 there is a service repository in every IEC 61499 resource for the FBIs to register their provided services, as shown in Fig.1 [7, Fig. 1]. This is performed by having each FBI to register the service definitions or service contracts, as claimed by authors. WSDL is used by authors in [7, Sec. V] to define service contracts, and the SOAP protocol is used to implement the interactions among FBIs in the same processing node.

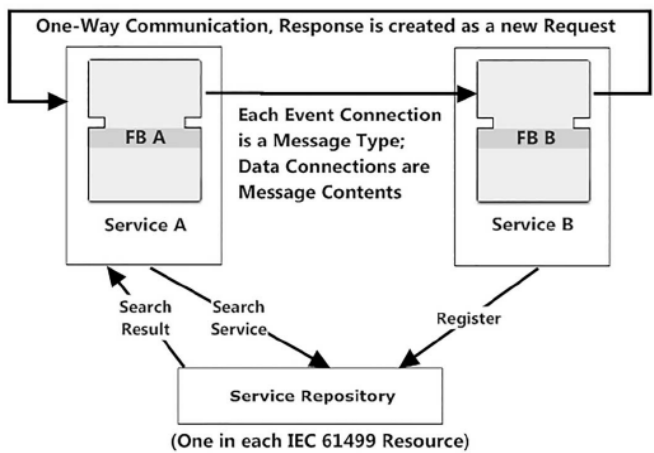

Fig. 1. The basic structure adopted in the formal SOA-based IEC 61499 model [7].

To the best of our knowledge this is the first attempt to utilize SOAP and WSDL to integrate the objects or components that constitute a controller software that is executed on one device. In [21] authors describe an approach for the integration of coordinate OO IEC 61131 FBs [27] with FBs that encapsulate plant resources, such as silos and pipes, adopting the event-based model of IEC 61499. They consider their approach as a service-oriented architecture. This approach is further discussed in [22].

It should be noted that basic FBs defined by the IEC 61499 standard include among others FBs for performing logic operations such as AND, OR, XOR as well as FB for merging (E_MERGE) and delaying (E_DELAY) of events. All these FBs are integrated based on the proposed approach using SOAP, WSDL and the WS-discovery protocol. FBIs register their services to the resource repository for other FBIs of the same device to discover and use these services. Part 3 of the IEC 61499 standard recommends practitioners to avoid using even CORBA with the argument that implementation of the features specified by its model would be too expensive, and its performance "would almost always be too slow, for use in a distributed real-time industrial-process measurement and control system (IPMCS)."

Authors with Definition 3 argue that services provided by composite FBs (CFBs) are considered as composite, based on the fact that CFBs are defined as a network of FBIs. Thus, a CFB is defined as an aggregation of services possibly using BPEL or a similar language for orchestrating smaller and fine-grained services provided by the CFB's constituent FBIs. The relation of this language with the definition process of Composite FB Types is not discussed.

Authors consider [7, Definition 5] the event and data connections among FBIs as one-way communication and consider response messages send by the provider FBI to be implemented by using a service that would be provided for this reason by the service requestor FBI. It is assumed that the motivation for this is Definition 4 and the graphical notation of the FBN which captures the response to a request as a separate event connection along with the corresponding data connections. This raises, among others, the question of how these two independent services will be combined in the service definition by using WSDL.

## IV. THE IEC 61499 SOA-BASED EXECUTION ENVIRONMENT ARCHITECTURE

### A. The Execution Environment

Authors in [7, Sec. V] describe an execution environment for IEC 61499 based on the formal model defined in the same paper. They present the key constructs of the execution environment using a class diagram [7, Fig.2] or [7, Fig.3] based on text [7, Sec IV]. From this diagram and Definition 1 that is utilized by authors to implement every class of this diagram as a service, it is extracted that the execution environment services, and the whole execution environment, are implemented using FB Types. The resource is implemented as a service repository but it keeps a list of FB types and FB instances. When a request for creating a new FB instance is received by the resource manager, one instance of the FB Service class is created and just one end point (the one of the FB Service instance) is registered in the repository of the resource, even in the case that the corresponding FB type provides more than one services that is the common case. The resource instance contains information not only on the provided by this instance services, but also for output events emitted by the FBI as well as data inputs and data outputs. This characterizes the resource repository as an FB instance repository and not service repository as claimed by authors.

From the definition of dynamic services it is extracted that not only input events are mapped to services but also the EC state algorithms. Data services are also defined to access internal variables of the FB instance. Service endpoints are also used for EC state actions, EC algorithms and EC actions and all these are stored in the service repository that means that SOAP and XML overhead is introduced even in the ECC execution time. Moreover, services are registered to the repository for every constituent FBIs of composite FB, that means that the overhead from service utilization is also introduced at the composite FB level. The WS-discovery protocol is utilized for service discovery from the resource repository. Even though the approach focus on distributed systems the relation of the resource repository with the device external one, that would probably be used to register device's exposed services is not discussed.

For the presented execution environment, authors assume that EC algorithms are normally written in IEC 61131 languages and mainly ST and LD. However, this raises the question of portability that was considered one of the main factors for the selection of 61499 instead of the 61131, which is claimed in [7] that does not provide code portability among various PLC vendors. On the other side it is claimed that code portability is achieved for FB library elements due to the use of their XML-based representation. It should be noted that PLCopen has defined an XML based representation for IEC 61131.

Authors claim [7, Sec.1] that interoperability can be achieved through the use to the publish/subscribe communication model. The use of publish/subscribe communication pattern for interaction assumes that publishers and subscribers have already addressed interoperability issues. The publish/subscribe communication pattern has been successfully utilized in IEC 61499 execution environments to obtain flexibility at the device level, e.g., [23][24].

*B. Run-time reconfiguration*

Run-time reconfiguration at the device level, which is considered as one benefit of the proposed architecture, imposes string real time constraints and complex algorithms not shown in [7]. The described case study even though considers the deletion and creation of FB types includes actions for deleting and creating event and data connections [7, Table I]. The creation of event connections among FBIs has to be related to the publish/discover based interaction on which the proposed architecture is based. The resource management model described by IEC 61499 to support the IDE in the deployment process is not consistent with the publish/discover model that authors have adopted for the construction of the formal model [7, Sec. IV]. For example, the management command of IEC 61499 "*CREATE event connection*" expresses a different model from the publish/discover pattern. A coordinator, the IDE, enforces the construction of an event connection among the specific FBIs. However, based on the publish/discover pattern and as authors claim, when an FBI "intends to invoke a particular logic from a service provider, the requested service will be located by the service repository for the service requester." Based on this, "the service requester can access the service provider via sending messages"

It is also interesting to note the feature of the framework that allows the control engineer to add a new functionality at the FB instance along with new services. This allows the control engineer, according to authors, to define FB types on the fly during normal operation and embed instances of them in the control logic.

Regarding the performance evaluation, the proposed execution environment is compared against FORTE, which is based on method call for FBI invocation. FORTE adopts a completely different execution semantics from the ones adopted in the proposed execution environment.

An overhead of 0.4 ms has been measured per persistent connection that is increased to 2.4 ms for temporary connections. It is clear that this last overhead has to be calculated for every connection of the new FB instance that is added to the network during reconfiguration at run-time. This probably results in more than 50 ms (this time is not reported in [7]) from the deletion of the old FB type till the end of the specific reconfiguration action described in the case study.

An execution environment for IEC 16499 that supports run time reconfiguration with detailed performance measurements is presented in [23]. Based on this: a) the average value of the FB instance creation time is 20 $\mu$s, and b) the creation of an event connection has an average time of 1.87 $\mu$s, while its deletion has an average value of 1.8 $\mu$s, both with a standard deviation of about 0.5 $\mu$s. It should be noted that RTNet is used as a communication mechanism instead of web services and SOAP.

SOAP has been developed to interconnect functionalities expressed in terms of software developed on heterogeneous hardware and/or software platforms, which are distributed over the internet. These two requirements, i.e., distribution and heterogeneity, do not exist in the single device IEC 61499 execution environment thus the cost of performance overhead and the complexity that its adoption introduces is without benefit.

## V. CONCLUSION

SOA has been evaluated by several research groups for its potential application in industrial automation systems. Research projects have resulted in the development of protocol stacks for the device level to allow the interconnection of the control PLCs with the upper layers of the manufacturing pyramid. However, SOAP and Web Services even though introduced in some PLCs have considerable performance overhead that is a big barrier in their use. The use of these technologies at the integration level of the device software constructs, greatly increases the performance overhead as well as the complexity at this level with questionable benefits regarding flexibility. Other technologies provide feasible solutions to this level of integration.

REFERENCES

[1] Jammes, F.; Smit, H., "Service-oriented paradigms in industrial automation", Industrial Informatics, IEEE Transactions on, Year: 2005, Volume: 1, Issue: 1, Pages: 62 - 70, DOI: 10.1109/TII.2005.844419

[2] Cucinotta, T. ; Mancina, A. ; Anastasi, G.F. ; Lipari, G. ; Mangeruca, L.; Checcozzo, R. ; Rusina, F. "A Real-Time Service-Oriented Architecture for *Industrial Automation"*, Industrial Informatics, IEEE Transactions on, Volume. 5, Issue. 3, August 2009, pp. 267-277.

[3] Candido, G.; Colombo, A.W.; Barata, J.; Jammes, F., "Service-Oriented Infrastructure to Support the Deployment of Evolvable Production Systems", Industrial Informatics, IEEE Transactions on, Year: 2011, Volume: 7, Issue: 4, Pages: 759 - 767, DOI: 10.1109/TII.2011.2166779

[4] Jestratjew, A.; Kwiecien, A., "Performance of HTTP Protocol in Networked Control Systems", Industrial Informatics, IEEE Transactions on, Year: 2013, Volume: 9, Issue: 1, Pages: 271 - 276, DOI: 10.1109/TII.2012.2183138

[5] K. Thramboulidis, D. Perdikis, S. Kantas, "Model Driven Development of Distributed Control Applications", The International Journal of

Advanced Manufacturing Technology, Volume 33, Numbers 3-4 / June, 2007, Springer-Verlag.
[6] Douglas S, Levine DL, Mungee S, "The design of the TAO real-time object request broker. Comput Comm 21(4), 1998, pp. 294–324.
[7] Dai, W.; Vyatkin, V.; Christensen, J.H.; Dubinin, V.N., "Bridging Service-Oriented Architecture and IEC 61499 for Flexibility and Interoperability", Industrial Informatics, IEEE Transactions on, Year: 2015, Volume: 11, Issue: 3 Pages: 771 - 781, DOI: 10.1109/TII.2015.2423495.
[8] M.P. Papazoglou, P. Traverso, S. Dustdar, F. Leymann, "Service-oriented computing: state of the art and research challenges", Computer, 40 (11) (2007), pp. 38–45
[9] W3C, SOAP Version 1.2 Part 1: Messaging Framework (Second Edition), Available on-line: http://www.w3.org/TR/soap12/
[10] Elfwing, R.; Paulsson, U.; Lundberg, L., "Performance of SOAP in Web Service environment compared to CORBA" Software Engineering Conference, 2002. Ninth Asia-Pacific
[11] Thramboulidis, K., "A cyber–physical system-based approach for industrial automation systems" Computers In Industry, Available on line 16 May 2015, doi:10.1016/j.compind.2015.04.006
[12] Li Da Xu, Wu He, and Shancang Li, "Internet of Things in Industries: A Survey", Industrial Informatics, IEEE Transactions on, Vol. 10, No. 4, November 2014
[13] The SIRENA project. [Online] Available: https://itea3.org/project/sirena.html
[14] The SODA project. [Online] Available: https://itea3.org/project/soda.html
[15] The SOCRADES project. [Online] Available: http://www.socrades.net/
[16] The SOA4D project. [Online] Available: https://forge.soa4d.org/softwaremap/trove_list.php
[17] Girbea, A.; Suciu, C.; Nechifor, S.; Sisak, F., "Design and Implementation of a Service-Oriented Architecture for the Optimization of Industrial Applications", Industrial Informatics, IEEE Transactions on, Year: 2014, Volume: 10, Issue: 1, Pages: 185 - 196, DOI: 10.1109/TII.2013.2253112
[18] Han, S.N.; Gyu Myoung Lee; Crespi, N. "Semantic Context-Aware Service Composition for Building Automation System", Industrial Informatics, IEEE Transactions on, Year: 2014, Volume: 10, Issue: 1, Pages: 752 - 761, DOI: 10.1109/TII.2013.2252356
[19] Puttonen, J.; Lobov, A.; Martinez Lastra, J.L., "Semantics-Based Composition of Factory Automation Processes Encapsulated by Web Services", Year: 2013, Volume: 9, Issue 4, Pages: 2349 - 2359, DOI: 10.1109/TII. 2012.2220554
[20] TwinCAT Beckoff, [Online] Available: http://www.beckhoff.com/english.asp?press/pr1414.htm
[21] F. Basile, P. Chiacchio, and D. Gerbasio, "On the Implementation of Industrial Automation Systems Based on PLC", IEEE Transactions on Automation Science and Engineering, Volume: 10, Issue: 4, pp.990-1003, Oct 2013.
[22] Thramboulidis, K., "A Cyber-Physical System-based Approach for Industrial Automation Systems", (submitted)
[23] G. Doukas, K. Thramboulidis, "A Real-Time Linux Based Framework for Model-Driven Engineering in Control and Automation", IEEE Transactions on Industrial Electronics, Vol. 58, No. 3, March 2011, pp. 914-924.
[24] K. Thramboulidis, S. Sierla, N. Papakonstantinou, K. Koskinen, "An IEC 61499 Based Approach for Distributed Batch Process Control", 5th IEEE International Conference on Industrial Informatics, July 23-27, 2007, Vienna, Austria.
[25] International Electrotechnical Commission, *International Standard IEC61499, Function Blocks, Part 1 - Part 4*, IEC, 2005.
[26] International Electrotechnical Commission, "IEC International Standard IEC 61131-3: Programmable Controllers, Part 3: Programming Languages," IEC, 2003.
[27] International Electrotechnical Commission, "IEC 61131-3 international standard", Edition 3.0 (2013-02-20) Programmable controllers - Part 3: Programming languages.
[28] W3C, "Simple Object Access Protocol (SOAP)", version 1.2, [Online] Available: http://www.w3.org/TR/soap12/
[29] OMG, Common Object Request Broker Architecture (CORBA), November 2012, [Online] Available: http://www.omg.org/spec/CORBA/3.3
[30] OASIS, Devices Profile for Web Services (DPWS), [Online] Available: http://docs.oasis-open.org/ws-dd/ns/dpws/2009/01